\documentclass[aps,prl,reprint,superscriptaddress,nofootinbib]{revtex4-2}

\usepackage[T1]{fontenc}
\usepackage[utf8]{inputenc}
\usepackage{amsmath,amssymb,amsfonts}
\usepackage{bm}
\usepackage{graphicx}
\usepackage[colorlinks=true,citecolor=blue,linkcolor=blue,urlcolor=blue]{hyperref}

\makeatletter
\newif\ifqrbmequalcontrib
\newcommand{\qrbmequalcontrib}{\global\qrbmequalcontribtrue}
\let\qrbm@affil@present@script\@affil@present@script
\renewcommand{\@affil@present@script}{%
  \qrbm@affil@present@script
  \ifqrbmequalcontrib
    \textsuperscript{,\,*}%
    \global\qrbmequalcontribfalse
  \fi
}
\makeatother

\newcommand{\SM}{SM~\cite{SupplementalMaterial}}

\begin{document}

\title{A Restricted Boltzmann Machine with Quantum-State Visible Units}

\author{Zhe-Hao Zhang}
\thanks{These authors contributed equally to this work.}
\affiliation{Institute for Advanced Study in Physics and School of Physics, Zhejiang University, Hangzhou 310027, China}

\author{Yi-Cong Yu\qrbmequalcontrib}
\email[Corresponding author: ]{ycyu@wipm.ac.cn}
\affiliation{Wuhan Institute of Physics and Mathematics, Innovation Academy for Precision Measurement Science and Technology, Chinese Academy of Sciences, Wuhan 430071, China}

\author{Xiaoming Cai}
\affiliation{Wuhan Institute of Physics and Mathematics, Innovation Academy for Precision Measurement Science and Technology, Chinese Academy of Sciences, Wuhan 430071, China}

\author{Hai-Qing Lin}
\email[Corresponding author: ]{hqlin@zju.edu.cn}
\affiliation{Institute for Advanced Study in Physics and School of Physics, Zhejiang University, Hangzhou 310027, China}

\date{\today}

\begin{abstract}
We construct a restricted Boltzmann machine (RBM) whose visible input is a quantum state rather than a classical configuration. Each hidden unit carries a trainable quantum template prepared by a parametrized circuit and converts its overlap with the input into a feature. Treating quantum states as high-dimensional continuous visible objects creates nontrivial normalization and scaling problems. We regularize the continuous likelihood and derive two controlled high-dimensional limits, yielding a Hopfield-type network with continuous Hilbert-space patterns and a data-augmented Gram likelihood. The resulting algorithms are compact and use trainable circuit-prepared templates as measurement intermediaries between quantum data and classical optimization. Numerical simulations across several many-body systems demonstrate effective quantum-phase recognition and multicomponent feature extraction.
\end{abstract}

\maketitle

\begin{figure}[t]
	\centering
	\includegraphics[width=\linewidth]{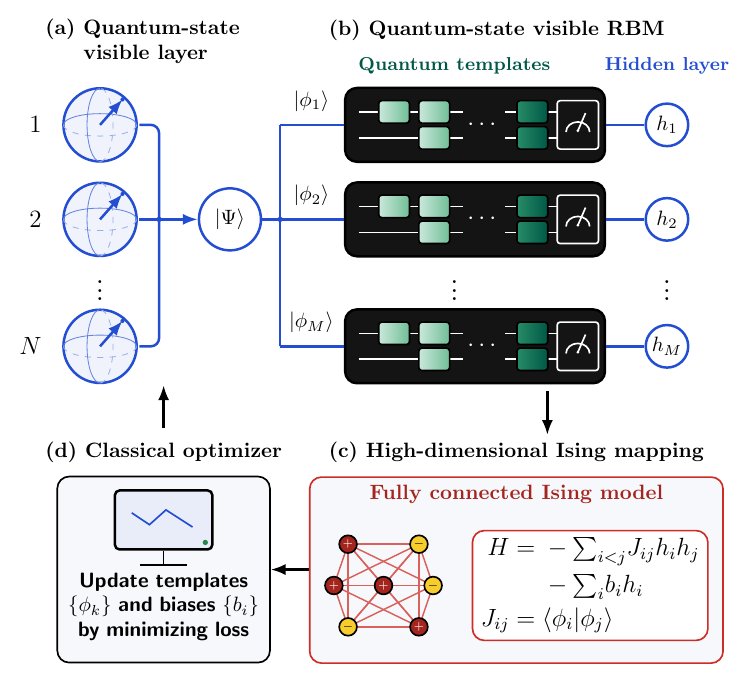}
	\caption{Schematic of the quantum-state visible RBM. (a) A normalized many-body state $|\Psi\rangle$ serves directly as the visible input. (b) Each binary hidden unit $h_j$ is associated with a normalized quantum template $|\hat\phi_j\rangle$, implemented by a parametrized circuit, together with a classical gain and hidden bias. (c) In the linear high-dimensional limit, integration over visible states yields a fully connected Ising model with weighted Gram couplings $J_{ij}=\bar\kappa_i\bar\kappa_j\langle\hat\phi_i|\hat\phi_j\rangle$. (d) A classical optimizer updates the circuit parameters, gains, and biases by minimizing the loss.
	}
	\label{fig0}
\end{figure}

\emph{Introduction.---}
The restricted Boltzmann machine (RBM) is a bipartite stochastic neural network that compresses visible data into learned hidden features. Its statistical-mechanical lineage runs from Hopfield associative memories~\cite{Hopfield1982,Hopfield1984} to stochastic Boltzmann machines with hidden degrees of freedom~\cite{HintonSejnowski1986}. The restricted architecture and contrastive divergence~\cite{Hinton2002} later supported early deep learning~\cite{Hinton2006DBN}, nonlinear dimensionality reduction~\cite{HintonSalakhutdinov2006}, and collaborative filtering~\cite{Salakhutdinov2007}.

RBMs entered quantum many-body physics through neural quantum states (NQS), which use an RBM to represent many-body wave functions~\cite{CarleoTroyer2017}. This variational program has expanded to correlated and symmetry-rich systems~\cite{Nomura2017,Lu2019,Vieijra2020}, developed a theory of expressivity, entanglement, sampling, and learnability~\cite{Deng2017,GaoDuan2017,Feng2026,DiSarra2026,Mummaneni2026}, and incorporated deeper architectures, improved optimization, transformers, and specialized hardware~\cite{ChenHeyl2024,Cao2024,Leone2026,Chowdhury2025}. In parallel, RBMs have been used for quantum-state tomography and learning from measurement data~\cite{Torlai2018,Beach2019,Zhao2023,Tonner2026}, while related work has quantized the Boltzmann architecture or introduced quantum-assisted training~\cite{Amin2018,Lyakhova2021,MoroPrati2023,Demidik2025,ElYazizi2025,Alsheikh2026}. These developments are reviewed in Refs.~\cite{Melko2019,CarrasquillaTorlai2021,Gebhart2023}.

In these quantum applications the RBM either consumes classically reduced data, as in neural-network quantum-state tomography and learning~\cite{Torlai2018,Beach2019,Huang2020,Zhao2023,Tonner2026}, or parameterizes the wave function, as in variational neural quantum states and their extensions~\cite{CarleoTroyer2017,Nomura2017,Lu2019,Vieijra2020,ChenHeyl2024,Leone2026}. The quantum state thus enters these schemes only as classical data or as classical parameters. We instead feed the state itself into the network as the visible variable, before any classical reduction, and let each hidden unit extract a feature through its overlap with a trainable quantum template, joining a growing line of work that studies quantum states directly~\cite{Lloyd2014,Schuld2019,Havlicek2019,JainKalev2025,SchuldPetruccione2021,Huang2022} [Fig.~\ref{fig0}]. With a quantum state as input, the discrete sum over $2^L$ configurations of the classical RBM becomes a continuous integral over Hilbert-space directions, making normalization and scaling dimension dependent. Treating these effects mathematically leads to new structures and training schemes.

\emph{Quantum-state visible RBM.---}
Throughout, we work with real state vectors in a fixed global-sign convention, which suffices for the real Hamiltonians considered below; the extension to complex states is discussed in the \SM{}.
For a classical RBM with visible vector $\bm v$, centered hidden spins $\bm h\in\{-1,1\}^{m}$, and connection matrix $\bm W=(\bm w_1,\ldots,\bm w_m)$, the energy is
\begin{equation}
    E_{\rm cl}(\bm v,\bm h)
    =-\sum_{j=1}^{m}h_j
    \left(\bm w_j^{\mathsf T}\bm v+b_j\right).
    \label{eq:classical-energy}
\end{equation}
Since $\bm W^{\mathsf T}$ is a learned linear map, its direct Hilbert-space extension is $\bm v\to|\Psi\rangle$, the input quantum state, and $\bm w_j\to|\phi_j\rangle$, the trainable template state, so that the projection $\bm w_j^{\mathsf T}\bm v$ becomes the matching signal $s(\Psi,\phi_j)$. For an $L$-site system, we parameterize each template by
\begin{equation}
    |\phi_j(\bm{\theta}_j)\rangle
    =
    \kappa_j\,|\hat\phi_j\rangle,
    \qquad
    |\hat\phi_j\rangle
    =
    U_j(\bm{\theta}_j)|0\rangle^{\otimes L}.
\end{equation}
Here $U_j(\bm{\theta}_j)$ is a unitary evolution operator, and all trainable quantum parameters reside in it, whereas $\kappa_j$ is a trainable classical scale; the physical input $|\Psi\rangle$ remains normalized, but $|\phi_j\rangle=\kappa_j|\hat\phi_j\rangle$ need not be. The matching signal $s$ is task dependent: it may be the sign-insensitive squared overlap $|\langle\Psi|\phi_j\rangle|^2$~\cite{Fuchs1999,Buhrman2001} or the signed linear overlap $\langle\Psi|\phi_j\rangle$~\cite{Kitaev1995,Ekert2002,NielsenChuang2010}. The quantum energy is then
\begin{equation}
    E(\Psi,h)
    =
    -\sum_{j=1}^{m} h_j
    \left[
        s(\Psi,\phi_j)+b_j
    \right],
    \label{eq:energy}
\end{equation}
which is Eq.~\eqref{eq:classical-energy} with each classical projection replaced by a quantum-state matching signal.

The high dimension is not innocuous. For generic normalized states in $d=2^L$ dimensions, the overlap of two independent random states concentrates around its mean, so the typical overlap is $\langle\Psi|\hat\phi_j\rangle=O(d^{-1/2})$~\cite{PopescuShortWinter2006,Vershynin2018}. This $\sqrt d$ scaling is familiar in machine learning: the scaled dot-product attention of formula (1) of Ref.~\cite{Vaswani2017} divides its logits by $\sqrt{d_k}$ for a related high-dimensional geometric reason. In classical arithmetic such a factor is a harmless numerical convenience, but in a quantum system the dimension $d$ is exponential, and this scale becomes a severe problem: the barren-plateau problem of variational quantum circuits~\cite{McClean2018,Cerezo2021,Holmes2022} may be viewed as one manifestation of related high-dimensional scaling, in which both the feature signal and its gradient can be suppressed. This random-state estimate identifies the natural asymptotic scale. Guided by it, we construct two controlled large-$d$ models with $\kappa_j=\sqrt d\,\bar\kappa_j$ and $\bar\kappa_j=O(1)$, rather than directly optimize the original finite-$d$ spherical RBM. The linear and squared readouts carry one and two powers of this scale, respectively, and yield the two effective training objectives below.

\emph{Likelihood on quantum-state space.---}
Unlike the classical spin RBM, whose visible space is the finite set $\Omega=\{0,1\}^{L}$, a normalized state in the real representation belongs to $\mathbb S^{d-1}\subset\mathbb R^d$, where $d$ is now the ambient dimension and the visible space is the continuous sphere. The sum over hidden configurations remains discrete, but the visible-layer sum becomes an integral with the normalized rotation-invariant measure $d\omega(\Psi)$ on $\mathbb S^{d-1}$ (defined in Sec.~I of the Supplemental Material (SM)~\cite{SupplementalMaterial}; see also Refs.~\cite{Fisher1953,Watson1983,MardiaJupp2000}).

For parameters $\bm{\theta}$ (the circuit angles, gains, and biases entering the quantum energy), define
\begin{equation}
    p_{\bm{\theta}}(\Psi)
    =
    \frac{1}{Z}
    \sum_{\bm{h}} e^{-E_{\bm{\theta}}(\Psi,\bm{h})},
    \label{eq:visible-density}
\end{equation}
where $E_{\bm\theta}(\Psi,\bm h)$ is the quantum energy of Eq.~(\ref{eq:energy}) with its parameters collected in $\bm\theta$, and $Z=\int_{\mathbb{S}^{d-1}}d\omega(\Psi)\sum_{\bm h}e^{-E_{\bm{\theta}}(\Psi,\bm h)}$. For training states $\{|\Psi_\ell\rangle\}_{\ell=1}^{M}$, the formal loss is $\mathcal{L}(\bm{\theta})=-M^{-1}\sum_{\ell}\log p_{\bm{\theta}}(\Psi_\ell)$. Although $p_{\bm\theta}$ is well defined at every finite parameter value, its finite-sample maximum-likelihood problem is ill posed. In the present RBM family, the unconstrained concentration scale generates normalized components of vanishing angular width around a training direction, while other hidden configurations retain a finite background. A peak of width $\varepsilon$ reaches height $O(\varepsilon^{-(d-1)})$; hence the empirical log likelihood is unbounded and has no finite maximizer~\cite{GoodGaskins1971,Silverman1982}. The explicit spherical construction is given in Sec.~I of the \SM{}.

We remove this singularity by regularizing the probability measure before optimization: specifically, we use spherical convolution, standard in directional statistics and spherical density estimation~\cite{Fisher1953,Bingham1974,Kent1982,Watson1983,MardiaJupp2000,HallWatsonCabrera1987}, with the rotationally invariant von Mises--Fisher kernel
\begin{equation}
    K_\lambda(\Psi,x)
    =Z_\lambda^{-1}e^{\lambda\langle\Psi|x\rangle},
\end{equation}
By rotational invariance, $Z_\lambda=\int_{\mathbb S^{d-1}}d\omega(y)e^{\lambda\langle\Psi|y\rangle}$ depends only on $\lambda$. The strong-quadratic branch instead uses $K_{\mathrm q}(\Psi,x)=C_d^{-1}e^{d|\langle\Psi|x\rangle|^2}$, normalized as specified in Sec.~IV of the \SM{}. The kernel-smoothed likelihood, written here for the linear branch, is
\begin{equation}
    \widetilde p_{\bm{\theta}}(\Psi_\ell)
    =
    \int_{\mathbb{S}^{d-1}} d\omega(x)\,
    K_\lambda(\Psi_\ell,x)p_{\bm{\theta}}(x).
    \label{eq:smoothed-likelihood}
\end{equation}

\emph{Controlled high-dimensional mappings.---}
Equation~(\ref{eq:smoothed-likelihood}) is well defined but intractable at large $d$; two controlled dimension-dependent limits yield closed effective training objectives.

\emph{(i) Linear-overlap limit.---}
For $\lambda=\sqrt d$, together with template scales
$|\phi_j\rangle=\sqrt d\,\bar\kappa_j|\hat\phi_j\rangle$, the spherical integral has a Gaussian limit and yields the Ising representation below, where the centered hidden configuration $\bm h$ is relabeled as $\bm\sigma$,
\begin{equation}
    Z_{\rm Ising}(\bm{b}_0,\bm{J})
    =
    \sum_{\bm{\sigma}}
    e^{
        \sum_j b_{0,j}\sigma_j
        +
        \sum_{i<j}J_{ij}\sigma_i\sigma_j
    },
    \label{eq:ising-partition}
\end{equation}
The coupling matrix and the data-induced field are defined separately by
\begin{equation}
    J_{ij}=\bar\kappa_i\bar\kappa_j\langle\hat\phi_i|\hat\phi_j\rangle,
    \qquad
    b_j^{(\ell)}=\bar\kappa_j\langle\Psi_\ell|\hat\phi_j\rangle .
    \label{eq:gram-coupling}
\end{equation}
For one datum the convolution shifts the base field $\bm b_0$ by $\bm b^{(\ell)}$. As derived in Secs.~I--III of the \SM{}, up to factors independent of the hidden configurations, the resulting optimization problem is
\begin{equation}
    \mathcal L_{\rm Hopfield}
    =-\frac{1}{M}\sum_{\ell=1}^{M}
    \log\frac{Z_{\rm Ising}(\bm b_0+\bm b^{(\ell)},\bm J)}
    {Z_{\rm Ising}(\bm b_0,\bm J)}.
    \label{eq:hopfield-loss}
\end{equation}

Equation~(\ref{eq:hopfield-loss}) has the Hopfield form because its coupling is Hebbian. In a Hilbert-space basis $\{|a\rangle\}$, define $\xi_j^{(a)}=\bar\kappa_j\langle a|\hat\phi_j\rangle$; Eq.~(\ref{eq:gram-coupling}) then gives $J_{ij}=\sum_a\xi_i^{(a)}\xi_j^{(a)}$, up to the conventional overall normalization. Conventional Hopfield patterns have binary components, whereas the components $\xi_j^{(a)}$ here are continuous. Integrating out the quantum-state visible layer therefore connects Hinton's RBM to a continuous-pattern generalization of Hopfield's associative-memory model~\cite{HintonSejnowski1986,Hinton2002,Hopfield1982,Hopfield1984}.

\emph{(ii) Strong-quadratic limit.---}
For the strong-quadratic branch considered here, we set $\bar\kappa_j=1$. The squared readout $s=|\langle\Psi|\phi_j\rangle|^2$ then carries the scale $d=(\sqrt d)^2$. 
In this regime the all-plus hidden configuration dominates, and the spherical integral is controlled by the direction that maximizes the resulting quadratic form. 
For $A=\sum_j|\hat\phi_j\rangle\langle\hat\phi_j|$ and $Z_{\mathrm{q}}=\int d\omega(x)e^{d\langle x|A|x\rangle}$, a saddle-point evaluation detailed in Sec.~IV of the \SM{} gives
\begin{equation}
    \frac{1}{d}\log Z_{\mathrm{q}}
    =\Phi_{\mathrm{q}}\!\left[\lambda_{\max}(\bm G)\right]+o(1),
    \label{eq:gram-saddle}
\end{equation}
where $\Phi_{\mathrm q}(\lambda)=\lambda-\frac12-\frac12\log(2\lambda)$ for $\lambda>1/2$. For a datum $|\Psi_\ell\rangle$, the relevant Gram matrices are
\[
    \begin{gathered}
        G_{ij}=\langle\hat\phi_i|\hat\phi_j\rangle,
        \qquad
        c_j^{(\ell)}=\langle\hat\phi_j|\Psi_\ell\rangle,\\
        \bm G^{(\ell)}
        =
        \begin{pmatrix}
            \bm G & \bm c^{(\ell)}\\
            (\bm c^{(\ell)})^{\mathsf T} & 1
        \end{pmatrix}.
    \end{gathered}
\]
Up to a template-independent constant, the dimension-scaled log likelihood is $d^{-1}\log\widetilde p_{\bm\theta}(\Psi_\ell)=\Phi_{\mathrm q}[\lambda_{\max}(\bm G^{(\ell)})]-\Phi_{\mathrm q}[\lambda_{\max}(\bm G)]+o(1)$. Accordingly, we minimize its limiting negative sum,
\begin{equation}
    \mathcal L_{\rm Gram}
    =\sum_{\ell=1}^{M}\left\{
    \Phi_{\mathrm q}
    \!\left[\lambda_{\max}(\bm G)\right]
    -
    \Phi_{\mathrm q}
    \!\left[\lambda_{\max}(\bm G^{(\ell)})\right]
    \right\}.
    \label{eq:gram-loss}
\end{equation}
Since $\bm G$ is a principal submatrix of each $\bm G^{(\ell)}$, Cauchy interlacing gives $\lambda_{\max}(\bm G^{(\ell)})\geq\lambda_{\max}(\bm G)$~\cite{HornJohnson2012}; monotonicity of $\Phi_{\mathrm q}$ therefore makes every term in Eq.~(\ref{eq:gram-loss}) nonpositive. Unlike the Hopfield mapping, this objective is governed by the leading eigenvalue of a quadratic overlap kernel. The resulting spectral-difference optimization is generally nonconvex and nonsmooth at leading-eigenvalue degeneracies~\cite{BoydVandenberghe2004,LewisOverton1996}.


\begin{figure}[t]
	\centering
	\includegraphics[width=\linewidth]{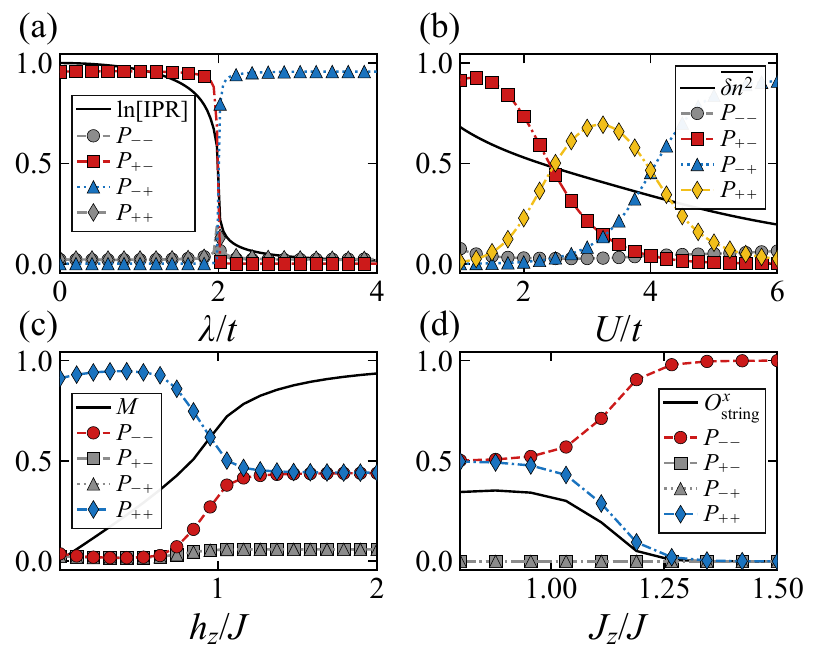}
	\caption{Phase classification by the minimal quantum-state visible RBM with two hidden units. Colored curves give the conditional probabilities $P_{\rm RBM}$ ($P_{--}$, $P_{+-}$, $P_{-+}$ and $P_{++}$), the black solid curve shows conventional physical indicators. (a) Aubry--Andr\'e model: inverse participation ratio rescaled across the localization transition at $\lambda/t=2$ ($L=377$ and $\beta=(\sqrt{5}-1)/2$). (b) Bose--Hubbard model at unit filling: averaged on-site number fluctuation across the superfluid--Mott crossover ($L=12$); the thermodynamic-limit critical value is $U/t\approx3.275$ \cite{thamm2025}. (c) Transverse-field Ising chain: magnetization across the symmetry-breaking transition near $h_z/J=1$ ($L=40$). (d) Spin-1 XXZ chain: transverse string order across the Haldane--N\'eel transition ($L=40$); the thermodynamic-limit critical value is $J_z/J\approx1.185$ \cite{chen2003}.
    }
	\label{fig1}
\end{figure}

Gram-matrix spectral methods are widely used in classical feature extraction~\cite{Scholkopf1998}, and overlap-matrix spectra are established many-body diagnostics~\cite{Zanardi2007,CamposVenuti2007}. Unlike these fixed-kernel constructions, and unlike qPCA~\cite{Lloyd2014}, our $\bm G$ is assembled from trainable template projectors and enlarged datum by datum through $\bm G^{(\ell)}$, making the RBM a trainable spectral analogue of qPCA with a quantum-kernel interpretation~\cite{Schuld2019,Havlicek2019}.

Both limits use parametrized circuits to probe inputs and optimize overlap features through their respective likelihoods. Product states are the simplest ansatz; the framework applies to any circuit-preparable template family.
\begin{figure}[t]
	\centering
	\includegraphics[width=\linewidth]{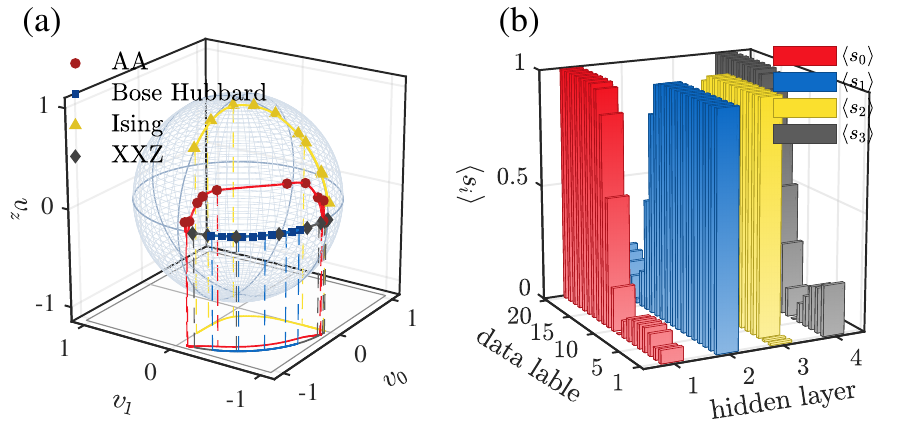}
	\caption{Quantum-state feature extraction. (a) Unit-sphere trajectories for the four benchmark data sets constructed from $v_0$, $v_1$, and $v_z$. (b) Conditional hidden-spin means $\langle\sigma_i\rangle$ for four learned templates and 20 transverse-field Ising ground states, ordered by their data labels. The distinct response profiles provide a multicomponent representation of the input states beyond phase classification.}
	\label{fig2}
\end{figure}

\emph{Results.---}
We first test the linear-overlap Hopfield construction on phase-label-free recognition, and then examine multicomponent feature extraction and the strong-quadratic Gram formulation. The unlabeled training set comprises ground states along Hamiltonian parameter sweeps, obtained by matrix-product-state methods~\cite{White1992,Schollwock2011} or exact diagonalization. Phase labels and transition locations are never used. The Hamiltonian-parameter order is used for visualization and, where specified in the \SM{}, to organize initialization and curriculum; the optimized objectives themselves depend only on state overlaps. Four representative one-dimensional models are studied: the Aubry--Andr\'e, unit-filled Bose--Hubbard, transverse-field Ising, and spin-1 XXZ chains.

For the MPS calculations we use bond-dimension-one product states $|\phi_j(\bm\theta_j)\rangle=\bigotimes_{n=1}^{L}(\cos\theta_{j,n}|0\rangle+\sin\theta_{j,n}|1\rangle)$, which are prepared by shallow circuits of local rotations. The dense exact-diagonalization benchmarks use normalized dense template directions as numerical controls. The classical gains and biases never enter the template circuits. For compactness, we drop the hat on normalized templates. The phase-boundary benchmarks use the minimal RBM with two hidden units, whose four joint configurations define four competing probability channels. The template choices and training protocols are detailed in Secs.~V and VI of the \SM{}.

Figure~\ref{fig1} compares the resulting $P_{\rm RBM}$ with a standard diagnostic for each model: the inverse participation ratio, local number fluctuation, magnetization, and transverse string order, respectively. In all four cases, the hidden-configuration distribution reorganizes in the same parameter region in which the corresponding physical diagnostic changes. The two-hidden-spin model therefore demonstrates that the constrained Hopfield objective in Eq.~(\ref{eq:hopfield-loss}) can extract characteristic state features across localization, interaction-driven, symmetry-breaking, and topological transitions. As a complementary geometric visualization, Fig.~\ref{fig2}(a) maps the two signed overlap channels onto the upper unit hemisphere by a radial projection. The plotted coordinates $(v_0,v_1)$ preserve the direction of the unprojected channel pair, while $v_z$ completes the unit vector. This visualization, defined precisely in Sec.~VI of the \SM{}, reveals characteristic state trajectories across the four transitions.


The two-hidden-spin model above yields four hidden-configuration probabilities, but some remain nearly featureless. We next extend the overlap-based construction to multicomponent feature extraction using a phase-insensitive repulsive squared-overlap coupling motivated by the Gram structure of Eq.~(\ref{eq:gram-coupling}), which suppresses redundant template responses. Figure~\ref{fig2}(b) shows that the four learned hidden units develop distinct response profiles across the input-state sequence. The distinction, annealed training, and temperature dependence are detailed in Sec.~VII of the \SM{}.

\begin{figure}[t]
    \centering
    \includegraphics[width=\linewidth]{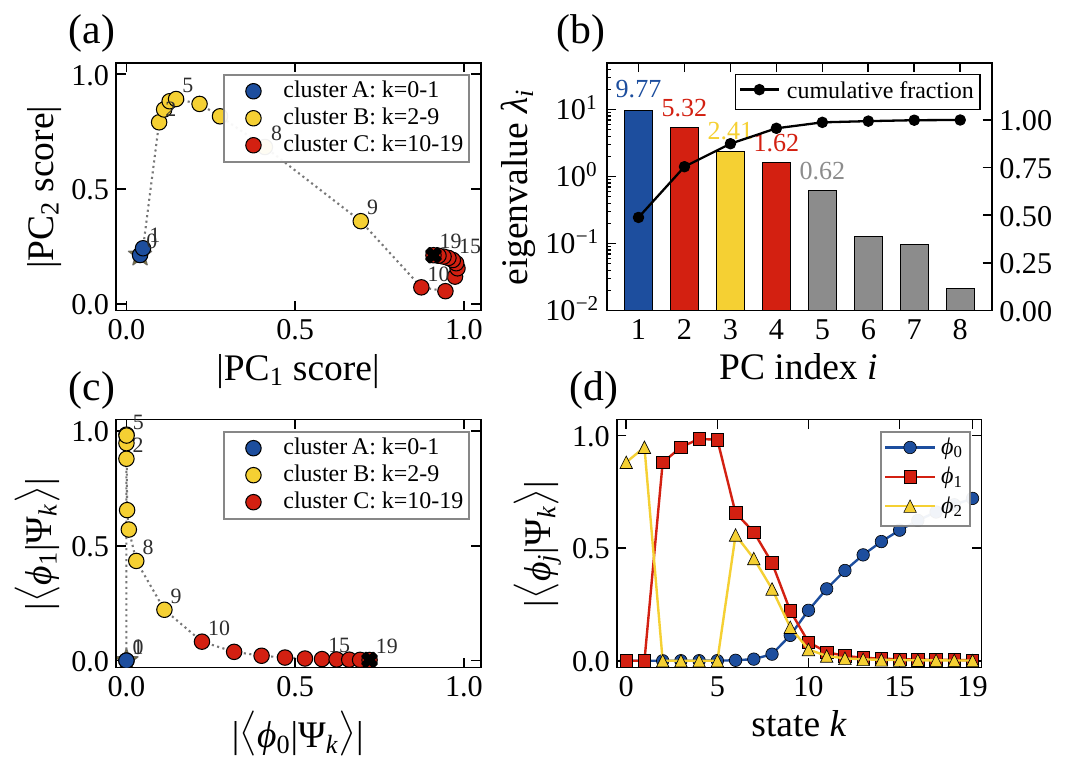}
    \caption{PCA and quantum-RBM representations of the same 20 transverse-field Ising states. (a) Absolute scores along the first two principal components; colors mark the three state regions $k=0$--$1$, $2$--$9$, and $10$--$19$. (b) Leading PCA eigenvalues and their cumulative fraction. (c) The same states represented by the first two of three learned overlap features. (d) All three features $|\langle\phi_j|\Psi_k\rangle|$.}
    \label{fig3}
\end{figure}

We next test the strong-quadratic Gram likelihood of Eq.~\ref{eq:gram-loss}. As a baseline, we apply kernel PCA~\cite{Scholkopf1998} to the input Gram matrix $K_{k\ell}=\langle\Psi_k|\Psi_\ell\rangle$, the quantum-state counterpart of unsupervised phase diagnostics~\cite{Wang2016,vanNieuwenburg2017}. The absolute scores along its first two principal directions trace three well-separated regions [Fig.~\ref{fig3}(a)], while the spectrum in Fig.~\ref{fig3}(b) shows that a few modes contain nearly all of the weight. We then optimize $P=3$ product-state templates with the Gram likelihood in Eq.~(\ref{eq:gram-loss}) and represent each datum by
\begin{equation}
    f_j(\Psi_k)=|\langle\phi_j|\Psi_k\rangle|.
    \label{eq:template-feature}
\end{equation}
The first two features expose the same three regions [Fig.~\ref{fig3}(c)], while the full profiles resolve the state sequence into distinct response channels [Fig.~\ref{fig3}(d)]. These regions correspond to the ferromagnetically ordered, critical, and field-polarized parts of the trajectory, so the learned features capture the underlying physics without input--input overlaps or phase labels. On the same 20-state, $L=40$ data set, the gain rises rapidly through $P=3$ and improves only weakly thereafter; as detailed in Sec.~VIII of the \SM{}, additional templates either remain redundant or refine an existing response region. Thus three broad feature profiles dominate this data set within the product-template family.

This kernel-PCA baseline requires $O(M^2)$ distinct input--input overlap observables. Our construction instead uses $P$ trainable templates and $O(PM)$ input--template overlap entries per objective evaluation at fixed $P$, without overlaps between distinct inputs. Arbitrary input--input overlaps generally require nonlocal or interferometric protocols~\cite{Buhrman2001}, whereas readout of the learned product-template features requires only local rotations and computational-basis measurement. Gram training additionally requires relative-sign information, which these fidelity measurements alone do not provide. Shot costs remain overlap dependent, so practical implementation requires a template family and initialization with resolvable support on the data manifold. The PCA comparison, additional measurement requirements, and Gram-training results are detailed in Sec.~VIII of the \SM{}.

\emph{Conclusion.---}
We have introduced an RBM whose visible data are quantum states rather than classical configurations or measurement records. Controlling the normalization and scale of the high-dimensional state space yields two effective training objectives: a Hopfield-type model and a strong-quadratic likelihood governed by data-augmented Gram eigenvalues. In both branches, trainable quantum circuits act as intermediaries: they probe each input through a small set of template overlaps, and the resulting features are optimized by the corresponding classical objective. The product-state circuits used here are a minimal realization, not a restriction of the framework. This circuit-mediated construction provides a new route to feature extraction from high-dimensional quantum states.

\emph{Acknowledgments.---}
We thank Sheng Yang for valuable discussions. 
This work was supported by MOST 2022YFA1402701, the National Natural Science Foundation of China (Grant No.~12274419), the Quantum Science and Technology--National Science and Technology Major Project (Grant No.~2023ZD0300404), and the CAS Project for Young Scientists in Basic Research (Grant No.~YSBR-055).

\emph{Data availability.---}
The computational codes and processed data used to generate the results of this work are publicly available at
\url{https://github.com/Mob-137/quantum-rbm-state-learning}.

\nocite{AubryAndre1980,Cristianini2002,DonohoElad2003,Fisher1989,Gentinetta2023,Haldane1983,Kerenidis2021,Kiefer1959,Miyato2018,Pfeuty1970,Zbontar2021}
\bibliographystyle{apsrev4-2}
\bibliography{refs}

\end{document}